# Using Quasigroups for Generating Pseudorandom Numbers

Vinod Kumar Godavarty

Email: vgodavarty@unomaha.edu

*Abstract -* This paper presents an algorithm for generating pseudorandom numbers using quasigroups. Random numbers have several applications in the area of secure communication. The proposed algorithm uses a matrix of size *n* x *n* which is pre-generated and stored. The quality of random numbers generated is compared with other pseudorandom number generator using Marsaglia's Diehard battery of tests.

## I. INTRODUCTION

Random numbers have applications in lottery games, shuffling cards, online gambling and secure communication. They may be used in generating random element for information dispersal [1] [2] and secret sharing algorithms [3]. Another application may be found in generation of shared secret keys between sensors in a sensor network [4] using factors of randomly generated matrices. In general, pseudorandom number generators use initial set of seeds to produce number sequences which have features close to random numbers i.e. without prior information about the seeds used the number sequence looks close to random numbers statistically [5]. A random number generator, with good statistical properties, based on a recursive algorithm is proposed in [6]. Several factors are involved in choosing a pseudorandom number generator i.e. memory constraints, computational power available etc. This paper presents pseudo random number generator based on Quasigroups which uses few resources.

Quasigroups or Latin Squares are *n* x *n* matrices similar to Sudoku. The computational cost of the scheme is very low, because it needs only table look up operations to create sequence of numbers and storage requirement for two *n* x *n* matrices. The final algorithm is tested for quality of randomness with Marsaglia's Diehard battery of statistical tests and test results are compared with other pseudorandom number generators.

## II. QUASIGROUPS

A Quasigroup is an *n* x *n* matrix consisting of permutation of elements of an finite field $Z_p$ such that no element repeats in any row or column and all elements appear in every row and column. Quasigroup satisfies the following laws,

$$\forall u, v \in Q, \exists x, y \in Q, \text{ such that, } u \cdot x = v \text{ and } y \cdot u = v$$
$$x \cdot y = x \cdot z \Rightarrow y = z \text{ and } y \cdot x = z \cdot x \Rightarrow y = z$$

Quasigroups also support adjoint operations denoted by / and \, and defined by,

$$x \cdot y = z \Leftrightarrow y = x \backslash z \Leftrightarrow x = z/y$$

Quasigroup operation denoted by $x \cdot y = z$ is a table lookup operation, where *z* is the element in row *x* and column *y* in Quasigroup.

| . | 1 | 2 | 3 | 4 | 5 |
|---|---|---|---|---|---|
| 1 | 2 | 1 | 5 | 3 | 4 |
| 2 | 5 | 4 | 2 | 1 | 3 |
| 3 | 3 | 5 | 1 | 4 | 2 |
| 4 | 4 | 2 | 3 | 5 | 1 |
| 5 | 1 | 3 | 4 | 2 | 5 |

Table 1: A quasigroup of order 5

Further the properties of Quasigroups or Latin Squares are described in [7].

Example: Consider Quasigroup shown in table 1. Let $u=2$ and $v=3$, then there exists $x$ and $y$ such that $u \cdot x = v$ and $y \cdot u = v$. Consequently, here $x=5$ and $y=5$.

| . | 1 | 2 | 3 | 4 | 5 |
|---|---|---|---|---|---|
| 1 | 2 | 1 | 5 | 3 | 4 |
| 2 | 5 | 4 | 2 | 1 | 3 |
| 3 | 3 | 5 | 1 | 4 | 2 |
| 4 | 4 | 2 | 3 | 5 | 1 |
| 5 | 1 | 3 | 4 | 2 | 5 |

Table 2: Property of quasigroup

### III. PROPOSED ALGORITHM FOR GENERATING PSEUDORANDOM NUMBERS

Let us consider Quasigroup '*qGroup*' which is an *n* x *n* matrix which is predetermined using randomly choosing the numbers from 1 to n and having it as matrix elements. The proposed generation is done in three phases. Phase1: A generator matrix, '*genMatrix*' is created from the previous generator matrix. In the first iteration, the generator matrix is created from the original quasigroup '*qGroup*'. Phase2: The elements of the *genMatrix* are read starting from first row and first column and moving through the columns till the end of the row and repeating the same for other rows. These read elements are sent out as pseudorandom numbers. Phase3: *genMatrix* is transposed and elements are shifted by a constant number *shiftConstant* or a *variableShift*. The *variableShift* can be defined as an element in *genMatrix* in specific row *x* and column *y* which are predefined. Algorithms for each phase are described below.

**Algorithm for Phase 1**:

1. if initialization=true
2.    *tempMatrix = qGroup*
3. else
4.    *tempMatrix = genMatrix*
5. end if
6. for i=1 to *order*
7. if i<order
8.    for j=1 to *order*
9.      if j <*order*
10.        *genMatrix* [i][j]= *tempMatrix* [i][j]· *tempMatrix* [i][j+1]
11.      else if j =order
12.        *genMatrix* [i][j]= *tempMatrix* [i][j]· *tempMatrix* [i+1][1]

13.      end if
14.     end for
15.    end if
16.   else if i = *order*
17.    for j=1 to *order*
18.     if j <*order*
19.      *genMatrix*[i][j]= *tempMatrix* [i][j]· *tempMatrix* [i][j+1]
20.     else if j =*order*
21.      *genMatrix*[i][j]= *tempMatrix* [i][j]· *tempMatrix* [1][1]
22.     end if
23.    end for
24.   end if
25. end for
26. initialization =false

"." Operation is look up operation on the Quasigroup matrix.

**Algorithm for Phase 2**: Reading the elements for pseudorandom numbers

1. for i=1 to *order*
2.  for j=1 to *order*
3.   read *genMatrix*[i][j]
4.  end for
5. end for

**Algorithm for Phase 3**:

1. Set *genMatrix*= transpose of *genMatrix*
2. List the rows of the matrix as elements of a single stream.
3. *variableShift = genMatrix*[*x*][*y*]; *x, y* are pre-defined
4. Rotate the stream by *shiftConstant* or *variableShift* of elements to right, based on choice.
5. Refill the elements from left in the *genMatrix* row-wise

**Algorithm for PRNG**:

1. initialization =true;
2. Execute Phase 1;
3. set initialization= false;
4. Execute phase 2
5. Execute phase 3
6. Repeat step 2, 3 and 4 until required length of random numbers are generated.

The idea behind this algorithm is to generate random numbers using quasigroup, until required length is reached. One of the advantages of this algorithm is that phase 2 is interlaced with phase 1, so we do not have to wait for complete generation of *genMatrix* and numbers can be read on fly as and when new elements of *genMatrix* are generated.

**Example for Proposed Algorithm**:

Let us consider the quasigroup 'Q' as the quasigroup in table 1. And consider the *shiftConstant* as 2.

*Iteration 0*: Phase 1: *tempMatrix* is same as original quasigroup '*qGroup*' and quasi operations are performed between adjacent elements to generate *genMatrix* i.e. *genMatrix* =

Step1: To find element in first row and first column of *genMatrix*, perform look up of element in first row and first column of *tempMatrix* and its adjacent element i.e. look up 2·1 in original quasigroup matrix

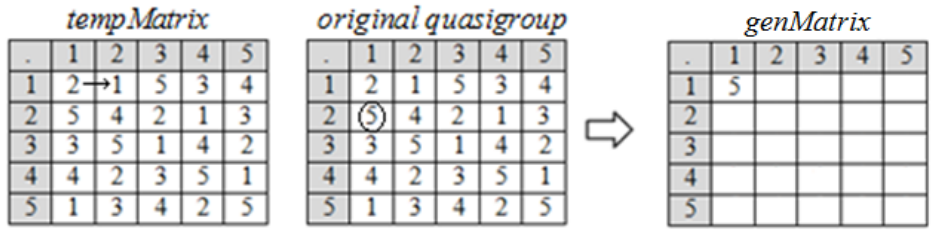

Step2: To find element in first row and second column of *genMatrix*, perform look up of element in first row and second column of *tempMatrix* and its adjacent element i.e. look up 1·5 in original quasigroup matrix

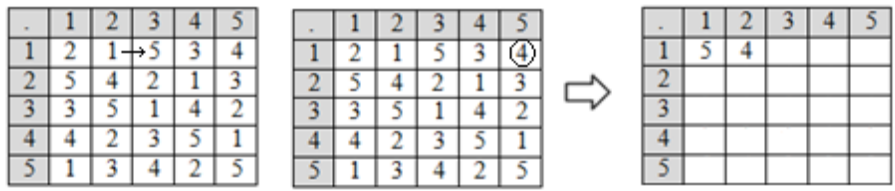

Further Steps: Perform similar operation as above for all other elements in *genMatrix*.

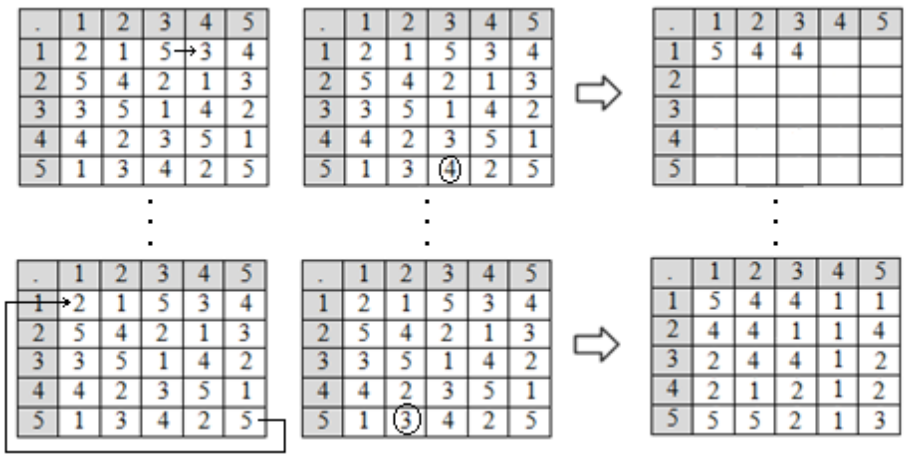

*Iteration 0*: Phase 2: Read the elements of *genMatrix* starting from the element in first row and column and move across the row. The number sequence generated serves as pseudorandom number sequence i.e. random numbers =

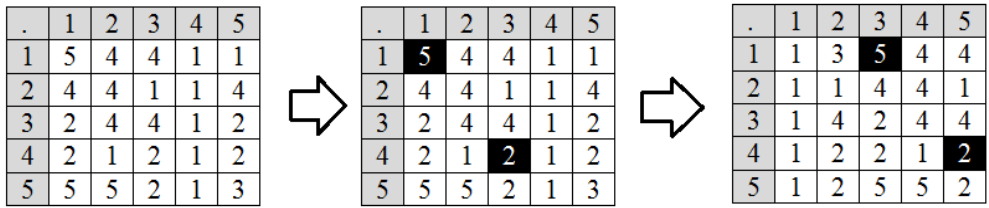

*Iteration* 0: Phase 3: Transpose the *genMatrix* and rotate the stream by *shiftConstant*= 2 i.e. *genMatrix* =

| . | 1 | 2 | 3 | 4 | 5 |
|---|---|---|---|---|---|
| 1 | 5 | 4 | 4 | 1 | 1 |
| 2 | 4 | 4 | 1 | 1 | 4 |
| 3 | 2 | 4 | 4 | 1 | 2 |
| 4 | 2 | 1 | 2 | 1 | 2 |
| 5 | 5 | 5 | 2 | 1 | 3 |

⇒

| . | 1 | 2 | 3 | 4 | 5 |
|---|---|---|---|---|---|
| 1 | 5 | 4 | 4 | 1 | 1 |
| 2 | 4 | 4 | 1 | 1 | 4 |
| 3 | 2 | 4 | 4 | 1 | 2 |
| 4 | 2 | 1 | 2 | 1 | 2 |
| 5 | 5 | 5 | 2 | 1 | 3 |

⇒

| . | 1 | 2 | 3 | 4 | 5 |
|---|---|---|---|---|---|
| 1 | 1 | 3 | 5 | 4 | 4 |
| 2 | 1 | 1 | 4 | 4 | 1 |
| 3 | 1 | 4 | 2 | 4 | 4 |
| 4 | 1 | 2 | 2 | 1 | 2 |
| 5 | 1 | 2 | 5 | 5 | 2 |

*Iteration* 1: Phase 1: *tempMatrix* is assigned with *genMatrix*. And rest of the process is same as iteration 0.

## IV. COMPARISON OF RANDOMNESS WITH WELL-KNOWN RNG

To compare the results of proposed pseudorandom number generator for 8 bit quasigroup (i.e. order of the quasigroup is 256 x 256), KISS random number generator is chosen. KISS (Keep It Simple Stupid) is originally proposed by G. Marsaglia and A. Zaman in 1993 and is an effective pseudo random number generator based on three separate generators, of different sorts, and requires 4 seeds [8].

The diehard battery of test results for both KISS Generator and Random number generator in Proposed Encryption Algorithm are tabulated below.

| Test Name | KISS (P-values) | Proposed RNG (P-Values) |
|---|---|---|
| Overlapping 5-Permutation Test 1 | 0.916<br>Chi-square for 99 degrees freedom = 118.974 | .999764<br>Chi-square for 99 degrees freedom = 155.881 |
| Overlapping 5-Permutation Test 2 | .339149<br>Chi-square for 99 degrees freedom = 92.638 | .443277<br>Chi-square for 99 degrees freedom = 96.350 |
| Binary Rank Test for 31 x 31 Matrices | .431276 | .866206 |
| Binary Rank Test for 32 x 32 Matrices | .649050 | .806213 |

Proposed RNG is having higher p-values when compared to the p-values of KISS generator which indicates that the proposed RNG algorithm is having high quality.

## V. DISCUSSION ON PROPOSED ALGORITHM

The above algorithm uses only look up operations, and hence the algorithm is computationally inexpensive. Also the only storage requirement that is required for the algorithm is to store original matrix and temporary matrix for producing pseudorandom numbers; the memory requirement complexity is described as follows. Consider that $n$ x $n$ be the order of the matrix where each element is of byte size. The storage requirement is given by $2n^2$ bytes. If $n$ is 256 then the total storage requirement is $2(256)^2 =$ 128 Kbytes. Hence the proposed algorithm is inexpensive in terms of memory requirements. The computational power required is to perform few table look up operations and few shift operations. Hence we can see that computational power required is very low. So in conclusion, we can say that the above algorithm can be used in resource constraint environments for generating pseudorandom numbers. Also, the proposed algorithm is having very good random number generator properties as mentioned in the previous section.

## VI. CONCLUSION

In this paper, we have proposed a new pseudorandom number generator algorithm which utilizes few resources like computational power and memory. Initial set of seeds i.e. original quasigroup and shift constant determines the upcoming sequence of the random numbers. Statistical results are presented to support the quality of the proposed pseudorandom number generator algorithm. Also this algorithm can be used to support secure communication or cryptography which necessitates high quality pseudorandom number generators.

## VII. ACKNOWLEDGEMENT

I thank Abhishek Tripathi for comments on initial draft of the paper.